\documentclass[12pt]{article}

\usepackage[pdftex]{graphicx}

\usepackage{epsfig,cite}

\usepackage {amsmath}

\usepackage{color}
\usepackage{rotating}
\usepackage{amssymb}

\usepackage{slashed}

\include{epsf}

\newcount\timecount

\newcount\hours \newcount\minutes  \newcount\temp \newcount\pmhours

\hours = \time

\divide\hours by 60

\temp = \hours

\multiply\temp by 60

\minutes = \time

\advance\minutes by -\temp

\def\hour{\the\hours}

\def\minute{\ifnum\minutes<10 0\the\minutes

            \else\the\minutes\fi}

\def\clock{

\ifnum\hours=0 12:\minute\ AM

\else\ifnum\hours<12 \hour:\minute\ AM

      \else\ifnum\hours=12 12:\minute\ PM

            \else\ifnum\hours>12

                 \pmhours=\hours

                 \advance\pmhours by -12

                 \the\pmhours:\minute\ PM

                 \fi

            \fi

      \fi

\fi

}

\def\monthname{\relax\ifcase\month 0/\or January\or February\or

   March\or April\or May\or June\or July\or August\or September\or

   October\or November\or December\else\number\month/\fi}

\def\bold#1{\setbox0=\hbox{$#1$}%

     \kern-.025em\copy0\kern-\wd0

     \kern.05em\copy0\kern-\wd0

     \kern-.025em\raise.0433em\box0 }

\def\beq{\begin{equation}}

\def\eeq{\end{equation}}

\def\ga{\mathrel{\raise.3ex\hbox{$>$\kern-.75em\lower1ex\hbox{$\sim$}}}}

\def\la{\mathrel{\raise.3ex\hbox{$<$\kern-.75em\lower1ex\hbox{$\sim$}}}}

\def\gev{{\rm \, Ge\kern-0.125em V}}

\def\tev{{\rm \, Te\kern-0.125em V}}

\def\gyr{{\rm \, G\kern-0.125em yr}}

\def\slash#1{\rlap{\hbox{$\mskip 1 mu /$}}#1}%

\def\gappeq{\mathrel{\rlap {\raise.5ex\hbox{$>$}}

{\lower.5ex\hbox{$\sim$}}}}

\def\lappeq{\mathrel{\rlap{\raise.5ex\hbox{$<$}}

{\lower.5ex\hbox{$\sim$}}}}

\def\Toprel#1\over#2{\mathrel{\mathop{#2}\limits^{#1}}}

\def\m12{m_{1\!/2}}

\def\bea{\begin{eqnarray}}

\def\eea{\end{eqnarray}}

\def\beqar{\begin{eqnarray}}

\def\eeqar{\end{eqnarray}}

\def\m{{\cal m}}

\begin{document}

\begin{titlepage}

\pagestyle{empty}

\baselineskip=21pt

%\rightline{\tt astro-ph/yymmnnn}

\rightline{KCL-PH-TH/2012-45, LCTS/2012-31, CERN-PH-TH/2012-310}

\vskip 1in

\begin{center}

{\large {\bf Prima Facie Evidence against Spin-Two Higgs Impostors}}

\end{center}

\begin{center}

\vskip 0.5in

 {\bf John~Ellis}$^{1,2}$,
 {\bf Ver\'onica Sanz}$^{2,3}$
and {\bf Tevong~You}$^{1}$

\vskip 0.2in

{\small {\it

$^1${Theoretical Particle Physics and Cosmology Group, Physics Department, \\
King's College London, London WC2R 2LS, UK}\\

$^2${TH Division, Physics Department, CERN, CH-1211 Geneva 23, Switzerland}\\

$^3${Department of Physics and Astronomy, York University, Toronto, ON, Canada M3J 1P3}\\

}}

\vskip 1in

{\bf Abstract}

\end{center}

\baselineskip=18pt \noindent

%%%%%%%%%%%%%%%%%%%%%%%%%%%%%%%%%%%%%%%%%%%%%%%%%

{
The new particle $X$ recently discovered by the ATLAS and CMS Collaborations is
widely expected to have spin zero, but this remains to be determined. The leading
alternative is that $X$ has spin two, presumably with graviton-like couplings. We show that
measurements of the $X$ particle to pairs of vector bosons constrain such scenarios.
In particular, a graviton-like Higgs impostor in scenarios with a warped extra dimension
of AdS type is {\it prima facie} excluded, principally because they predict too small a
ratio between the $X$ couplings to $WW$ and $ZZ$, compared with that to photons. 
The data also disfavour universal couplings to pairs of photons and gluons, which would
be predicted in a large class of graviton-like models. 
}

%%%%%%%%%%%%%%%%%%%%%%%%%%%%%%%%%%%%%%%%%%%%%%%%

\vfill

\leftline{%CERN-PH-TH/2011-xxx, KCL-PH-TH/2011-xxx, LCTS-2011-yy, 
November 2012}

\end{titlepage}

\baselineskip=18pt

%%%%%%%%%%%%%%%%%%%%%%%%%%%%%%%%%%%%%%%%%%%%%%%%%%

\section{Introduction and Summary}

The ATLAS~\cite{ATLASICHEP2012} and CMS~\cite{CMSICHEP2012}
Collaborations have discovered a new particle $X$ with mass $\sim 125$ to 126~GeV
during their searches for the Higgs boson at the LHC. Supporting evidence for $X$
production in association with massive vector bosons $V \equiv W, Z$ at the TeVatron
has been provided by the CDF and D0 Collaborations~\cite{TevatronJulySearch}.
If it is indeed a/the Higgs boson of the Standard Model, the $X$ particle must have spin zero.
Since it has been observed to decay into pairs of photons, we already know that the $X$ 
particle cannot have spin one, but spin two is still an open possibility at the time of writing.

In view of the importance of determining the `Higgs' spin, and the strong presumption that
it has spin zero, it is particularly important to take an unbiased, approach to its measurement.
Indeed, there is an extensive literature on possible strategies to distinguish the spin-parity $J^P$
of the $X$ particle, based on the kinematic characteristics of its production and decays~\cite{spinstudies,EFHSY}.
Examples include correlations between the momenta of particles produced in $X$ decays into $\gamma \gamma$,
$WW^\ast$ and $ZZ^\ast$, and the $V+X$ invariant mass when it is produced in association with a massive vector boson $V$~\cite{EHSY}. 
It is generally expected that significant evidence on the possible spin
of the $X$ particle will shortly be provided by analyses of the existing TeVatron and 2012 LHC data.

In this paper we explore the extent to which the available data on $X$ production and decay already
provide {\it prima facie} evidence that it is not a spin-two particle with graviton-like couplings in the
frameworks of some popular models~\footnote{The
spirit of this analysis is similar to that of~\cite{CKL}, where {\it prima facie} evidence was presented that
the `Higgs' particle is not a pseudoscalar.}. As we recall in Section~2, the couplings $c_{g, \gamma}$ 
of a Higgs impostor $X$ to gluon pairs and photon pairs must be equal in many models
with a compactified extra dimension, and hence
\begin{equation}
\Gamma (X \to gg) \; = \; 8 \, \Gamma (X \to \gamma \gamma) \, . 
\label{ratio8}
\end{equation}
This relation is completely different from the case of a Higgs-like spin-zero particle, for which the
$Xgg$ and $X\gamma\gamma$ couplings are induced by loop diagrams, and
$\Gamma (X \to gg) = {\cal O}(\alpha_s/\alpha_{EM})^2 \, \Gamma (X \to \gamma \gamma)$.
Numerically, at the one-loop level for the Higgs boson $H$ in the Standard Model in the limit
$m_H \ll 2 \, m_t, 2 \, m_W$ one has
\begin{equation}
\Gamma (H \to gg) \; \simeq 37 \; \Gamma (H \to \gamma \gamma) \, .
\label{ratioH}
\end{equation}
Various analyses have shown that the current data are compatible with the $X$ particle being
a Standard Model Higgs boson~\cite{EY2, postdiscovery}, and in particular with (\ref{ratioH}).

Here we argue that the present data on $X$ production and decay disfavour the
graviton-like spin-two prediction (\ref{ratio8}), providing some {\it prima facie} evidence against the
spin-two hypothesis. However, some graviton-like spin-two interpretations of the $X$ particle
encounter more serious problems. For example, in models with a warped fifth dimension of AdS
type one expects the following hierarchy of couplings to the energy-momentum tensors of different
particle species:
\begin{equation}
c_b \; \simeq \; c_t \; \; \ga \; \; c_W \; \simeq \; c_Z  \; = \; {\cal O}(35) \times ( c_g \; = \; c_\gamma \; \; > \; \; c_u, \;  c_d) \, .
\label{chierarchy}
\end{equation}
As we show later, the hierarchy between $c_{W,Z}$ and $c_{g, \gamma}$ predicted in (\ref{chierarchy})
is in strong tension with the available data, which indicate a much greater hierarchy.

In the rest of this paper, we first review in Section~2 the couplings of a graviton-like spin-two boson,
emphasizing the model-independence of the prediction (\ref{ratio8}) and discussing the
motivations for the more model-dependent predictions (\ref{chierarchy}). We then discuss the
current data in Section~3, and the problems they raise for the predictions (\ref{ratio8}) and
(\ref{chierarchy}). Finally, in Section~4 we summarize our conclusions and discuss the prospects
for gaining further insight into the nature of the $X$ particle.

%%%%%%%%%%%%%%VERO
%%%%%%%%%%%%%%%%%%%%%%%%%

\section{Spin-Two Boson Couplings to Standard Model Particles}

It was pointed out in~\cite{us-G} that dimension-four couplings of a massive spin-two
particle to a pair of Standard Model particles are forbidden by Lorentz invariance and gauge symmetry.
The flavour and CP symmetries of the Standard Model then imply that the leading dimension-five terms
should be proportional to their energy-momentum tensors $T^{i}_{\mu \nu}$, so that the couplings take the forms
\begin{equation}
{\cal L}_{int}=-\frac{c_{i}}{M_{eff}} G^{\mu \nu} T^{i}_{\mu \nu} \, . 
\label{LG}
\end{equation}
In scenarios with extra dimensions, $M_{eff} \simeq {\cal O}$(TeV)
is the effective Planck mass, whereas in composite models $M_{eff}$
would be a scale related to confinement. These two scenarios are, in general, related
by some suitable extension of the AdS/CFT correspondence, and we consider here
the formulation in terms of an extra dimension.

We consider general warped geometries of the form 
 \bea
ds^2 \; = \; w^2(z) \, (\eta_{\mu \nu} d x^{\mu} d x^{\nu}-dz^2) \ ,  \label{geom}
\eea 
where $w(z)=1$ for a flat extra dimension, and in the case of warping {\it \`a la} AdS
one has $w(z) = z_{UV}/z$. In general, $w(z)$ is a positive constant or decreasing function of $z$.
 
In such a scenario, the Kaluza-Klein (KK) decomposition for spin-one particles leads to an
equation of motion for the wave-function  of the $n$th KK mode, $f_n(z)$, of the following form~\cite{not-summing}:
\bea
\partial_z \left(  w(z) \partial_z f_n(z)\right) \; = \; -m_n^2 w(z) f_n(z) \, .
\label{gaugeEOM}
\eea
If the four-dimensional gauge symmetry is preserved by the compactification,
as is the case for the SU(3) of QCD and the U(1) of electromagnetism, then
the spin-one field has a massless zero mode, i.e., the lowest-lying KK mode has $m_0 = 0$,
implying
\bea
w(z) \partial_z f_0(z) \; = \; \textrm{constant} \, .
\label{zeroEOM}
\eea
Taking into account the Neumann boundary conditions on the boundary branes, there is only one solution, namely 
\bea
f_0(z)=C \, ,
\eea 
where the constant $C$ is determined by requiring the canonical normalization for the
four-dimensional gauge field. 

Obviously, the graviton is not the source of electroweak symmetry breaking (EWSB). 
Instead, one may think that EWSB is triggered by a condensate of new fermions induced by new
strongly-interactinggauge fields, as in technicolor models~\cite{TC}, a heavy Higgs, or, in the 
language of models with extra dimensions, by boundary conditions~\cite{Higgsless}. 
The graviton would couple to this source of EWSB, which we can parametrize by a field $\Sigma$
that could be spurious or dynamical and satisfies $\langle\Sigma\rangle = v$. In view of
the small values of the $T$ and $\Delta \rho$ parameters~\cite{pdg}, the field
$\Sigma$ should respect an approximate custodial symmetry, and couple to the graviton via an
effective interaction of the form
\bea
\frac{c_{\Sigma}}{M_{eff}} \, G_{\mu\nu} D^{\mu} \Sigma D^{\nu} \Sigma \, ,
\eea
where
\bea
D^{\mu} \equiv \partial^{\mu} + i g W^{\mu}+ i g B^{\mu} \, .
\label{inherit-c}
\eea
Gauge invariance implies that the graviton couples to the gauge eigenstates $W^a$ universally in (\ref{LG}), 
i.e.,  $c_W = c_Z$ as $g'\to 0$. Once EWSB occurs, the graviton would feel the effect through couplings induced
via (\ref{inherit-c}), which also respect custodial symmetry. 

The next issue is the relation between $c_{\gamma,g}$ and $c_{W,Z}$.  If it is assumed that 
electroweak symmetry is broken by boundary conditions on the IR brane, the support of the
wave-functions of the transverse components of the $W$ and $Z$ is suppressed near
this brane, so that $c_{W_t, Z_t} < c_{g, \gamma}$. However, the wave-functions of the
longitudinal components of the $W$ and $Z$ are localized near the IR brane, as are
the wave-functions of the massive fermions $b$ and $t$, so that $c_{W_L, Z_L, b, t} > c_{g, \gamma}$. 
On the other hand, the wave-functions of light fermions such as the $u$ and $d$ are expected to to be concentrated 
closer to the UV brane, so that $c_{u,d} \ll c_{g, \gamma}$.

One can estimate the hierarchy between $c_{\gamma,g}$ and $c_{W,Z}$ by accounting for  the 
suppression due to the difference between localization on the IR brane, 
where the graviton has most of its support, and delocalization in the bulk. 
The couplings of the massive graviton to the massless gauge bosons,
i.e., the gluons and photon, are suppressed by the effective
volume of the extra dimension, namely~\cite{gap-metrics}
\beq
c_{g, \gamma} \; \simeq \; 1/\int_{z_{UV}}^{z_{IR}} w(z) dz \, ,
\label{cint}
\eeq
and are therefore universal, leading to the result (\ref{ratio8}).
If the extra dimension is of AdS type, $w(z)=1/k z$, and the suppression is
by a factor $\log{M_{Pl}/TeV}\simeq$ 35. In other metrics, one could get a different degree of suppression. 
For example, one could introduce deviations from conformal invariance in AdS 
(or condensates of canonical dimension $d$ in the dual picture), by introducing metrics of the form~\cite{non-ads}
\bea
w(z) = \frac{1}{k z} \, \left(1+ c_d \left(\frac{z}{z_{IR}}\right)^{2 d} \right) \, .
\label{quasi-ads}
\eea
Those effects would not change the AdS result
\bea
c_{W,Z}/c_{\gamma,g} \lesssim {\cal O} (35)
\eea
by more than a factor ${\cal O}(1)$. On the other hand, one could obtain a larger difference
by postulating a metric that is not asymptotically AdS. 

In the dual picture, metrics of the form (\ref{quasi-ads}) correspond to theories which become scale-invariant at high energies.  
This is a very attractive feature of a strongly-coupled theory, as one can relate low-energy quantities to the UV behaviour by 
using, for example, the operator product expansion~\cite{interpolating}. Therefore, $c_{W,Z}/c_{\gamma,g} \gg 35$ would 
mean that the composite theory does not have such behaviour in the UV, implying a loss of predictivity. 

This derivation was made assuming that QCD and electromagnetism are present in the bulk.
As an alternative, one could imagine localizing electromagnetism and strong interactions on
a brane located at $z_\ast \in (z_{UV},z_{IR})$, in which case
\beq
c_{g, \gamma} \; \simeq \; \frac{w(z_{IR})}{w(z_{\ast})} \, .
\label{astUVratio}
\eeq
leading again to the relation (\ref{ratio8}).
One could also imagine a  a situation where the gluon (or the photon) is stuck on a brane and the 
photon (or gluon) is in the bulk or on the opposite brane. However, this option
is phenomenologically very disfavoured, since quarks are charged 
under both gauge groups and would need a non-negligible overlap with both fields, possibly leading to a 
large five-dimensional gauge coupling $g_{5D}$, implying
a low cutoff of the effective theory~\cite{NDA-xdims}, as $\Lambda_{NDA}\propto 1/g_{5D}^2$.  Also note that in the dual picture, the spin-two resonance could be made up of states with no color or no electric charge and this would invalidate the relation between decays to photons and gluons. 

Localized kinetic terms do not modify this relation for massless gauge fields,
since they modify only the right-hand side of the equation of motion (\ref{gaugeEOM}),
which is proportional to the mass. In the case of a massless zero mode, the effect is on the 
normalization of the mode in the bulk, namely the relation between the five-dimensional 
gauge couplings $g_5$ and their four-dimensional equivalents. However, this effect is absorbed
by fixing the four-dimensional couplings of the zero-mode gauge fields to the Standard Model
values and re-scaling the KK couplings. Since the graviton decays to photons and gluons depend
only on the number of degrees of freedom, the relation (\ref{ratio8}) is unchanged.

To summarize, graviton-like couplings satisfy the following properties
\begin{itemize}
  \item Due to current conservation: $c_{g}= c_{\gamma}$ \, ,
  \item Custodial symmetry: $c_{W}=c_{Z}$ \, ,
  \item For the theory to be asymptotically scale-invariant: $c_{W,Z}  \lesssim {\cal O} (35) c_{\gamma,g}$ \, ,
\end{itemize}
which we exploit in the next Section of this paper. 

Before closing this Section, however, we should mention one observation that disfavours a graviton-like explanation for the `Higgs'. 
The observed state is very light, with a mass $\simeq 125$~GeV whereas, if the graviton is a manifestation of extra
dimensions, one would expect that the mass of the massive graviton would be given by
\bea
m_{G} \sim 1/z_{IR} \, .
\eea
As other fields also live in the extra dimension, one would also expect comparatively light excitations of these fields, 
with masses typically of the order of $1/z_{IR} \simeq m_h$. For example, in the minimal AdS case
$m_{s=2} \simeq 1.5 m_{s=1}$, which is clearly ruled out by direct constraints. For example, a $Z'$
resonance with mass of order 100  GeV and electroweak couplings is ruled out by TeVatron and LHC searches~\cite{zprime}. 
On the dual side, one would argue along the same lines, but with $1/z_{IR}$ being
replaced by the scale of confinement. In a QCD-like theory, one would expect the masses of the 
resonances to increase with the spin, so that the lightest tensor meson would be heavier than the 
vector analogue of the $\rho$ meson.
One might hope for a separation between the spin-two and spin-one excitations in metrics
of more general form than AdS, but this is not the case. Using the techniques in Ref.~\cite{not-summing}, one can show that
\bea
m^{-2}_{s} \simeq \int d z w(z)^{2 s-1} \int_z d y/ w(y)^{2 s-1} \ ,
\eea
where $s$ is the spin.
As $w(z)$ is a decreasing function, this leads to the conclusion that $m_{s=2} > m_{s=1}$.

Despite this objection, we consider the spin-two interpretation of the $X$ particle with an open mind,
guided by the expectations (\ref{chierarchy}).

\section{Interpretation of Experimental Measurements}

A generic experimental measurement of the number of $X$ particle events in any specific channel
is proportional to a quantity of the form $\Gamma_i \Gamma_f/\Gamma_\text{Tot}$, where
$\Gamma_f$ is the decay rate into the observed final state, $\Gamma_\text{Tot}$ is the total
decay rate, and $\Gamma_i$ is the rate at which the $X$ particle decays into the pair of
partons that produce it. In the Standard Model, the dominant production process is $gg \to X$
and $\Gamma_i$ represents the decay rate for $X \to gg$~\footnote{There are important QCD radiative
corrections in both the production cross section and the $gg$ decay rate, but these are similar in magnitude,
so their net effects are not important for our purposes. There are no such corrections in the graviton-like spin-two case,
since it couples to the energy-momentum tensor.},
though this should not be taken for granted in graviton-like models, as we discuss below. 
In the case of `Higgs'-strahlung in association with a vector boson $V = W$ or $Z$,
or of vector-boson fusion (VBF), $\Gamma_i$ represents the decay rate for $X \to V V^\ast$. 
In the Standard Model, processes with initial-state ${\bar b} b$ are negligible, in particular, because of the
small density of $b$ partons in the incident protons, but this also needs to be reviewed in
graviton-like models in view of the possibility that $c_b$ is enhanced as in (\ref{chierarchy}).
Likewise, processes with ${\bar t} t$ in the final state are
known to contribute $\la 0.5$\% of Higgs production in the Standard Model,
but could be more important in the graviton-like case. On the other hand,
we can neglect processes with initial-state ${\bar u} u$
and ${\bar d} d$ because their couplings to the $X$ particle are expected to be very small in both the
Higgs and graviton scenarios.

\begin{figure}
\vskip 0.5in
%\vspace*{-0.75in}
%\hspace*{-.70in}
\begin{minipage}{8in}
\hspace*{-0.7in}
\centerline{\includegraphics[height=8cm]{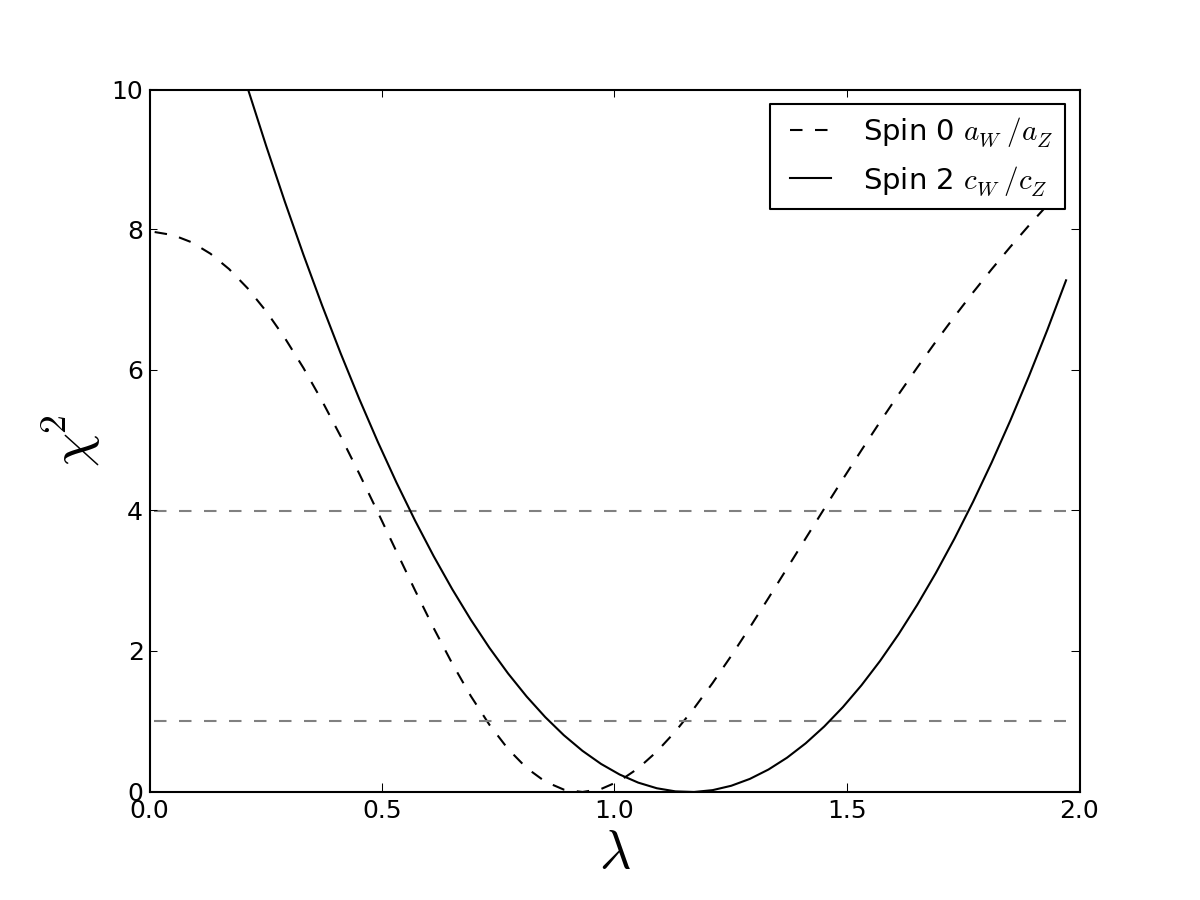}}
\hfill
\end{minipage}
\caption{
{\it
The constraints on the ratios $\lambda \equiv c_W/c_Z$ in the spin-two case
(solid line) and $a_W/a_Z$ in the spin-zero case (dashed line). Custodial
symmetry: $\lambda = 1$ is compatible with the data in both cases~\protect\cite{EFHSY}.}} 
\label{fig:Lambda}
\end{figure}

We first consider the experimental constraints on the ratios $\Gamma_\gamma/\Gamma_W$ and $\Gamma_\gamma/\Gamma_Z$.
Information on $\Gamma_\gamma/\Gamma_{W,Z}$ is provided by data on the $X$ branching ratios,
with only mild assumptions on the $X$ production mechanism(s) and spin. Concretely, we may write
\begin{equation}
\left. \frac{\Gamma_\gamma}{\Gamma_W}\right|_{2} \; = \; \frac{K_\gamma}{K_W} \left. \frac{\Gamma_\gamma}{\Gamma_W}\right|_0 \, ,
\label{gammaW20}
\end{equation}
and similarly for the $Z$ case,
where the notation $|_s$ denotes the spin hypothesis under which the decay modes are analyzed, and
the factors $K_f$ encode the differences in the kinematic acceptances for spin-two and spin-zero $X \to \gamma \gamma, WW^\ast$.
Based on~\cite{EFHSY}, we estimate that $K_\gamma \simeq 1$ (reflecting the large angular acceptances of
ATLAS and CMS for the $\gamma \gamma$ final state) whereas $K_W \simeq 1.9$ (reflecting the fact that the
ATLAS and CMS $WW^\ast$ analyses were optimized for the spin-zero hypothesis and have lower efficiencies
in the spin-2 case), so that
\begin{equation}
\left. \frac{\Gamma_\gamma}{\Gamma_W}\right|_2 \; = \; \frac{1}{1.9} \left. \frac{\Gamma_\gamma}{\Gamma_W}\right|_0 \, ,
\label{gammaW20K}
\end{equation}
The ratio $\Gamma_\gamma/\Gamma_W|_0$ may be parametrized in the form
\begin{equation}
\left. \frac{\Gamma_\gamma}{\Gamma_W}\right|_0 \; = \; \frac{\mu^{LHC}_{gg\to h \to\gamma\gamma}}{\mu^{LHC}_{gg\to h \to WW}} \times \left. \frac{\Gamma_\gamma}{\Gamma_W}\right|_{SM} \, , 
\label{gammaW0exp}
\end{equation}
where the signal strength factors $\mu^{LHC}_f$ are determined experimentally to be~\footnote{It is well-known that the measured values of
$\Gamma_\gamma/\Gamma_W$ and $\Gamma_\gamma/\Gamma_Z$ are somewhat higher than
expected in the Standard Model, due to the apparent enhancement of $X \to \gamma \gamma$ events, but
this does not have a big effect on our analysis.}:
\begin{eqnarray}
\mu^{LHC}_{gg\to h \to \gamma\gamma} &=& 1.54 \pm 0.28	\, , \\
\mu^{LHC}_{gg\to h \to WW} &=& 0.83 \pm 0.29	\, ,
\label{eq:muV}
\end{eqnarray}
We conclude that
\begin{equation}
\left. \frac{\Gamma_\gamma}{\Gamma_W}\right|_2 \; = \; (0.98 \pm 0.38) \times \left. \frac{\Gamma_\gamma}{\Gamma_W}\right|_{SM} \, .
\label{gammaW2SM}
\end{equation}
In the case of the $ZZ^\ast$ final state, we combine the ATLAS result with the CMS result obtained
without applying the MELA analysis~\cite{CMSICHEP2012}, obtaining
\begin{equation}
\mu^{LHC}_{gg\to h \to ZZ} \; = \; 0.91 \pm 0.25	\, .
\label{eq:muZ}
\end{equation}
Since we do not use the MELA analysis, there is no efficiency correction analogous to (\ref{gammaW20K}),
so we conclude also that
\begin{equation}
\left. \frac{\Gamma_\gamma}{\Gamma_Z}\right|_2 \; = \; (0.91 \pm 0.25) \times \left. \frac{\Gamma_\gamma}{\Gamma_Z}\right|_{SM} \, .
\label{gammaZ2SM}
\end{equation}
We have used {\tt MadGraph5}~\cite{MG5} to evaluate the decay rates for graviton-like $X \to WW^\ast$ and
$ZZ^\ast$ decay as functions of $c_{W,Z}/M_{eff}$ (\ref{LG}) and compared them with standard
calculations of graviton-like $X \to \gamma \gamma$ decay. Using a numerical computation, we estimate that
\begin{eqnarray}
\left. \frac{\Gamma_\gamma}{\Gamma_W}\right|_\text{Graviton} & \simeq & 280 \left(\frac{c_\gamma}{c_W}\right)^2 \, , \label{eq:NWNZg} \\
\left. \frac{\Gamma_\gamma}{\Gamma_Z}\right|_\text{Graviton} & \simeq & 2900 \left(\frac{c_\gamma}{c_Z}\right)^2 \, .
\label{eq:NWNZ}
\end{eqnarray}
Using (\ref{eq:NWNZ}), and the Standard Model
values $\Gamma_\gamma/\Gamma_W|_{SM} = 0.0106$ and $\Gamma_\gamma/\Gamma_Z|_{SM} = 0.086$, we find that the result (\ref{gammaW2SM}) corresponds to
\begin{eqnarray}
\frac{c_\gamma}{c_W} & = & 0.0061 \pm 0.0012 \, , \nonumber \\
\frac{c_\gamma}{c_Z} & = & 0.0071 \pm 0.0012 \, ,
\label{eq:cWcZ}
\end{eqnarray}
which are consistent with custodial symmetry:
\begin{equation}
\lambda \; \equiv \; \frac{c_W}{c_Z} \; = \; 1.16 \pm 0.30 \, ,
\label{custodial}
\end{equation}
as already shown in~\cite{EFHSY}. 
Fig.~\ref{fig:Lambda} compares the constraint
(\ref{custodial}) on the possible deviation of $\lambda \equiv c_W/c_Z$ from custodial 
symmetry in the spin-two case (solid line) with the corresponding ratio $a_W/a_Z$ in
the spin-zero case (dashed line), as shown in Fig.~15 of~\cite{EFHSY}.
We see that, though the spin-zero case gives a marginally better fit to the data, {\it there
is currently no significant preference over the spin two option}.

On the other hand, combining the results (\ref{eq:cWcZ}), we infer that
\begin{equation}
c_V \; = \; (175 \pm 25) \times c_\gamma \; .
\label{eq:cV}
\end{equation}
Qualitatively, this result is $\gg 1$, as is the graviton-like spin-two expectation (\ref{chierarchy}).
However, quantitatively the hierarchy (\ref{eq:cV}) is much larger than the ratio ${\cal O}(35)$
expected in AdS warped compactifications. {\it These models are disfavoured by well over
$3 \sigma$} and, as argued in Section~2, a larger value would be a sign of a theory which is not scale invariant in the UV.

In order to test whether $c_g = c_\gamma$, we consider the ratio of the rates for $gg \to X \to \gamma \gamma$ and
VBF $\to X \to \gamma \gamma$, which is related to $\Gamma_g/\Gamma_W$:
\begin{equation}
\frac{\Gamma_g}{\Gamma_W} \; = \; F_{0, 2} \;
\frac{\left[ \frac{\Gamma_g\Gamma_\gamma}{\Gamma_\text{Tot}}\right]^\text{LHC}_{gg\to h \to \gamma\gamma}}{\left[ \frac{\Gamma_W\Gamma_\gamma}{\Gamma_\text{Tot}}\right]^\text{LHC}_{VBF \to h \to \gamma\gamma}}	\quad , 
\label{eq:ggVBF}
\end{equation}
where $F_{0,2}$ are the ratios of initial flux factors, kinematic factors, efficiency factors, etc. appearing in 
the measured rates for the $gg$- and VBF-induced cross sections
under the spin-zero and graviton-like spin-two hypotheses, respectively. Assuming that these factors are similar
(differences by ${\cal O}(1)$ are unimportant), we may write
 \begin{equation}
\left. \frac{\Gamma_g}{\Gamma_W}\right|_\text{measured} / \left. \frac{\Gamma_g}{\Gamma_W}\right|_\text{SM} \; = \; 
\left. \frac{\left[ \frac{\Gamma_g\Gamma_\gamma}{\Gamma_\text{Tot}}\right]^\text{LHC}_{gg\to h \to \gamma\gamma}}{\left[ \frac{\Gamma_W\Gamma_\gamma}{\Gamma_\text{Tot}}\right]^\text{LHC}_{VBF \to h \to \gamma\gamma}}\right|_\text{measured} /
\left. \frac{\left[ \frac{\Gamma_g\Gamma_\gamma}{\Gamma_\text{Tot}}\right]^\text{LHC}_{gg\to h \to \gamma\gamma}}{\left[ \frac{\Gamma_W\Gamma_\gamma}{\Gamma_\text{Tot}}\right]^\text{LHC}_{VBF \to h \to \gamma\gamma}}\right|_\text{SM}	\quad , 
\label{eq:ggVBFmeas0}
\end{equation}
In order to evaluate this ratio, we use the parametrizations:
\begin{eqnarray}
\left. \left[ \frac{\Gamma_g\Gamma_\gamma}{\Gamma_\text{Tot}}\right]^\text{LHC}_{gg \to h \to \gamma \gamma} \right|_\text{measured} & = & \mu^\text{LHC}_{gg \to h \to \gamma \gamma} \times \left. \left[ \frac{\Gamma_g\Gamma_\gamma}{\Gamma_\text{Tot}}\right]^\text{LHC}_{gg \to h \to \gamma \gamma} \right|_\text{SM} \, , \nonumber \\
\left. \left[ \frac{\Gamma_W\Gamma_\gamma}{\Gamma_\text{Tot}}\right]^\text{LHC}_{VBF \to h \to \gamma\gamma} \right|_\text{measured} &= & \mu^\text{LHC}_{VBF \to h \to \gamma\gamma} \times \left. \left[ \frac{\Gamma_W\Gamma_\gamma}{\Gamma_\text{Tot}}\right]^\text{LHC}_{VBF \to h \to \gamma\gamma} \right|_\text{SM} \, ,
\label{eq:measSM}
\end{eqnarray}
and the following experimental values for the signal strength factors $\mu_i^\text{LHC}$:
\begin{eqnarray}
\mu^{LHC}_{gg\to h \to \gamma\gamma} &=& 1.54 \pm 0.28	\, , \\
\mu^{LHC}_{VBF \to h \to \gamma \gamma} &=& 1.98 \pm 0.84	\, ,
\label{eq:mugamma}
\end{eqnarray}
with the result
\begin{equation}
\left. \frac{\Gamma_g}{\Gamma_W}\right|_\text{measured} \; = \; (0.78 \pm 0.36) \times \left. \frac{\Gamma_g}{\Gamma_W}\right|_\text{SM} \, .
\label{eq:GammagW}
\end{equation}
where $\Gamma_g/\Gamma_W|_\text{SM} = 0.40 \, (0.36)$ for $M_h = 125 \, (126)$~GeV.
We also recall the Standard Model prediction $\Gamma_g/\Gamma_\gamma|_\text{SM} = 37$ for either value
of $M_h$. Combining this with (\ref{eq:GammagW}) and (\ref{gammaW2SM}),
we obtain the estimate
\begin{equation}
\left. \frac{\Gamma_g}{\Gamma_\gamma} \right|_\text{measured,2} \; \simeq \; 29 \pm 13 \, ,
\label{eq:measratio}
\end{equation}
where we have emphasized that this estimate applies to a spin-two graviton-like particle.
This estimate is to be compared with the prediction of graviton-like models:
\begin{equation}
\left. \frac{\Gamma_g}{\Gamma_W}\right|_\text{Graviton} \;  = \; 8 \left. \frac{\Gamma_\gamma}{\Gamma_W}\right|_\text{Graviton} \, ,
\label{badnews}
\end{equation}
which, we recall, is not subject to radiative corrections because the coupling is proportional to the energy-momentum tensor.
{\it There is clearly some tension between the data (\ref{eq:measratio}) and the prediction
(\ref{badnews}) of the graviton-like model.}

This discrepancy may be phrased in terms of the coefficient $c_g$ using (\ref{badnews})
in conjunction with the calculation (\ref{eq:NWNZg}):
\begin{equation}
\left. \frac{\Gamma_g}{\Gamma_W}\right|_\text{Graviton} \; \simeq \; 2200 \left(\frac{c_g}{c_W}\right)^2 \, .
\label{eq:NWNg}
\end{equation}
Putting (\ref{eq:GammagW}) and (\ref{eq:NWNg}) together, we find
\begin{equation}
\frac{c_g}{c_W} \; \simeq \; 0.012 \pm 0.0027 \, .
\end{equation}
Combining with (\ref{eq:cWcZ}), we infer that
\begin{equation}
c_g \; = \; (1.97 \pm 0.59) \times c_\gamma \; ,
\label{eq:cg}
\end{equation}
{\it in poor agreement with the graviton-like spin-two expectation (\ref{chierarchy})}.

\begin{figure}
\vskip 0.5in
\vspace*{-0.75in}
%\hspace*{-.70in}
\begin{minipage}{8in}
\hspace*{-0.7in}
\centerline{
{\includegraphics[height=6cm]{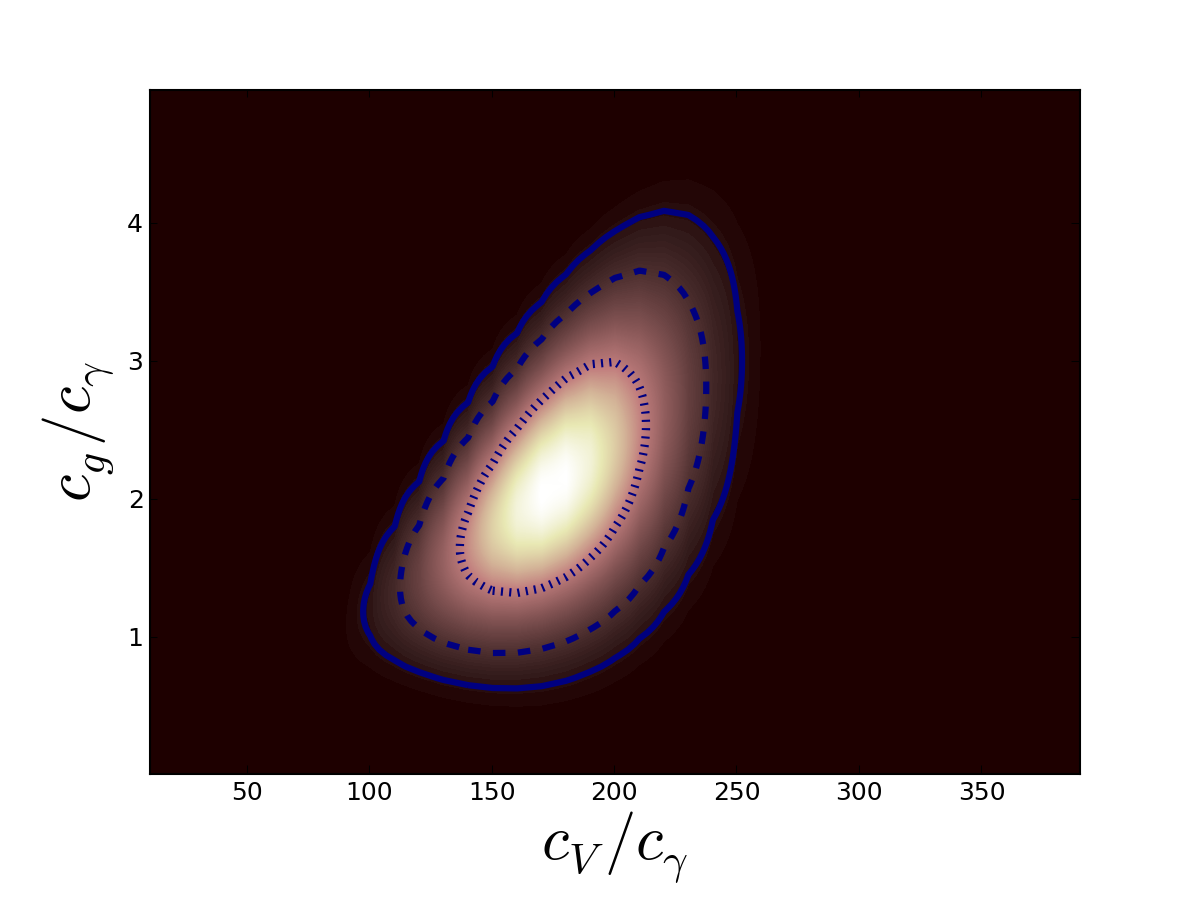}}
{\includegraphics[height=6cm]{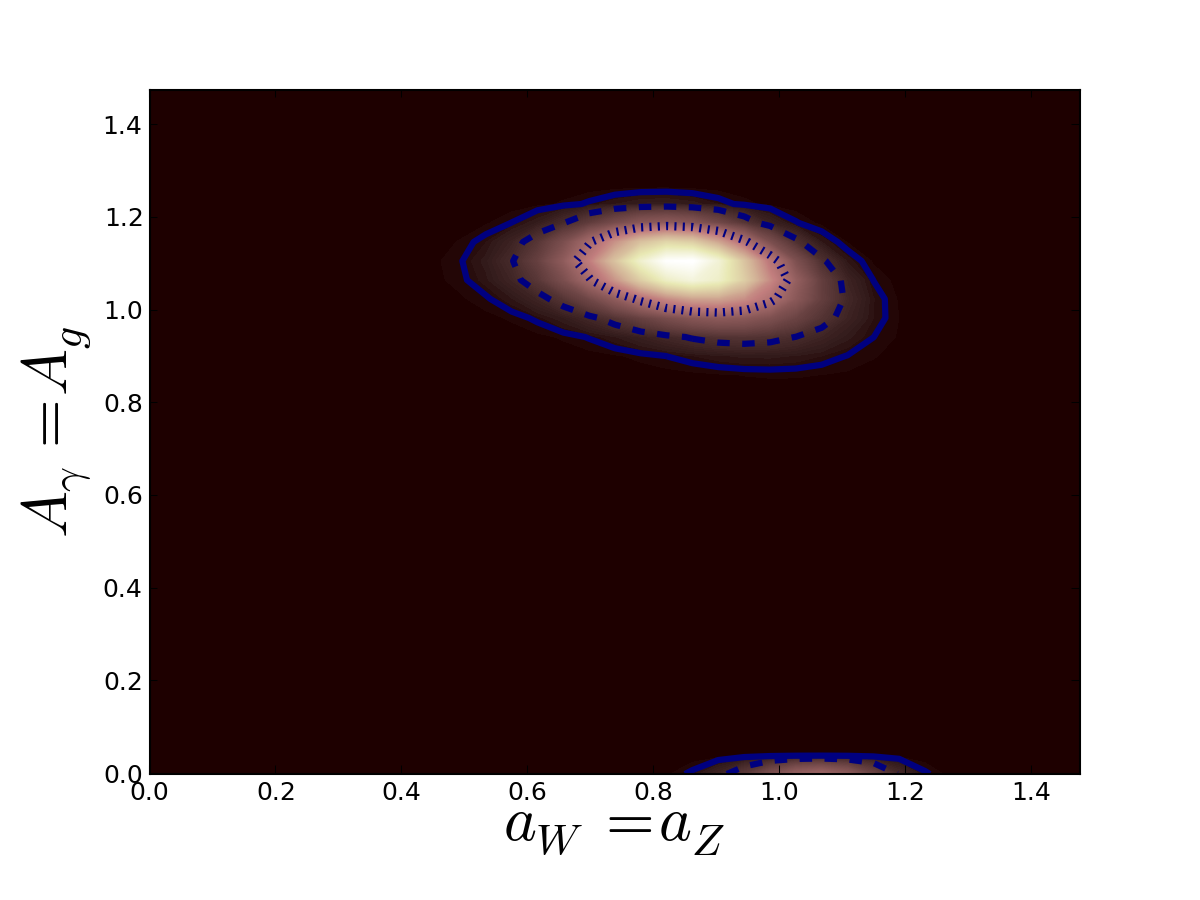}}
}
\hfill
\end{minipage}
\caption{
{\it
Left panel: the correlation between the values of $c_W/c_\gamma$ (horizontal axis) and $c_g/c_\gamma$
(vertical axis) found in a global fit to the current experimental data under the spin-two hypothesis. 
Right panel: a global fit under the spin-zero hypothesis to the couplings to massive vector bosons
(assuming custodial symmetry and a common ratio to the Standard Model values) and to massless 
vector bosons $g, \gamma$ (assuming a common ratio $A \equiv A_g = A_\gamma$ to the values of the
loop diagrams in the Standard Model).
}}  
\label{fig:planes} 
\end{figure}

The above results for graviton-like models are displayed in the left panel of Fig.~\ref{fig:planes},
which displays the correlation between the current
experimental constraints on $c_V/c_\gamma$ (horizontal axis) and $c_g/c_\gamma$
(vertical axis). We see that the best fit corresponds to $c_W/c_\gamma \sim 175$ as
shown in (\ref{eq:cV}) and $c_g/c_\gamma \sim 2$ as shown in (\ref{eq:cg}). We also
see a tendency for smaller values of $c_W/c_\gamma$ to be correlated with smaller
values of $c_g/c_\gamma$. However, we also see that the predictions
$(c_W/c_\gamma, c_g/c_\gamma) \sim (35, 1)$ of a graviton-like spin-two particle in
warped space-time are jointly disfavoured by $\gg 3 \sigma$. In contrast, we see in the right
panel of Fig.~\ref{fig:planes} that a global fit to all the available data under the
spin-zero hypothesis is very compatible not only with the couplings to massive vector
bosons having the Standard Model values (assuming custodial symmetry $a_W = a_Z$),
but also the triangle diagrams responsible for the 
couplings of a spin-zero particle to $gg$ and $\gamma \gamma$. Defining $A_{\gamma, g}$
to be the ratios of these triangle diagrams to their values in the Standard Model, we see that
the data are very compatible with them having a common value $A \equiv A_\gamma = A_g$
close to unity.

However, there is a potential loophole in the discussion of $c_g$. As recorded in (\ref{chierarchy}),
the coupling of a graviton-like spin-two particle to ${\bar b}b$ could be enhanced.
One may ask whether an enhanced $X {\bar b}b$ coupling could lead to $X$ production
processes involving $b$ quarks becoming sufficiently important to invalidate the above
argument. In particular, one should consider the possibility that they could contribute to
the event categories assumed by the ATLAS and CMS experiments to be due to $gg$
collisions, in which case (\ref{eq:GammagW}) would become an upper limit and,
accordingly, (\ref{eq:measratio}, \ref{eq:cV}) might be brought into agreement
with the universality prediction of the graviton-like model.

In the Standard Model, the parton-level cross section for $gg \to H$ is a factor $\sim 50$
smaller than the parton-level cross section for ${\bar b} b \to H$ in the four-flavour renormalization scheme.
Nevertheless, the total $gg \to H$ cross section ($\simeq 15$~pb at 7~TeV, see Table~1 of~\cite{HXSWG}) is much larger than the
total ${\bar b} b$-related cross section as calculated in either the four- or five-flavour scheme ($\simeq 250$~pb, see Figs.~22 and 23
of~\cite{HXSWG}). This is because the $gg$ parton collision luminosity factor is much larger than the corresponding factor
for ${\bar b} b$ collisions. In order to rescue the hypothesis that $c_g = c_\gamma$ in the graviton-like spin-two model, one would need
the total ${\bar b} b$-related cross section to exceed  total $gg \to H$ cross section by a factor $\sim 3$,
which would require $c_b/c_\gamma \simeq 100$. 

Such an enhancement is consistent, {\it a priori},
with the generic expectations (\ref{chierarchy}), and could be probed experimentally by
studying whether many ${\bar b} b$ pairs are produced in association with the $X$ boson. 
However, it would lead to suppressions of the {\it branching ratios}
for the decay modes $X \to \gamma \gamma, WW^\ast$ and $Z Z^\ast$.
On the other hand, the Fermilab observation of $X$ production in association with $W, Z$
implies that the {\it decay rates} for the decays $X \to WW^\ast$ and $Z Z^\ast$ must be
similar to their Standard Model values and hence, by extension, also the decay rate for $X \to \gamma \gamma$. 
Thus, the total $X$ decay width should be enhanced relative to its Standard Model value by a factor $\sim 10^4/(0.61/2 \times 10^{-3})
\sim 30$, so that $\Gamma_\text{Tot} \sim 100$~MeV: this is not inconsistent with the data.

As also seen in (\ref{chierarchy}), one would expect that $c_t \ga c_b$, with custodial symmetry
suggesting that $c_t \simeq c_b$. In this case, one would have $c_t/c_W \sim 100 \times (c_\gamma/c_W) \simeq 0.6$.
In this case, in the absence of a detailed calculation, one would expect the total
${\bar t} t X$ cross section to be smaller than in the Standard Model. We note that
CMS and ATLAS have currently established upper limits on this cross section that are 4.6 and 13.1 times
larger than the Standard Model prediction, respectively~\cite{ttbarH}, under the assumption that the $X$ particle has spin zero.
These upper limits need to be recalculated for the spin-two case, but there is no {\it prima facie}
contradiction with the possibility that $c_b/c_\gamma \simeq c_t/c_\gamma \simeq 100$. 

\section{Summary}

We have shown that the available data on $X$ production and decay already disfavour the
possibility that it is a spin-two impostor. It has been argued previously that such impostors
should have graviton-like couplings to other particles, and one expects $c_g = c_\gamma$
in all such scenarios. In the favoured warped compactifications of AdS type, one also expects
custodial symmetry so that $c_W = c_Z \equiv c_V$, and that $c_V/c_\gamma \simeq 35$. We have shown
that, whereas $c_W = c_V$ is compatible with the data, they favour $c_g > c_\gamma$ and
$c_V/c_\gamma \gg 35$. This last result is the strongest element in our {\it prima facie} case
against the $X$ particle being a spin-two Higgs impostor.

The advent of more data from the LHC 2012 run and more refined analyses of the TeVatron
data will enable our arguments to be sharpened. We also expect that the LHC and TeVatron
experiments will come forward with other, more direct, information about the spin of the $X$
particle. Probably nobody, least of all the authors, seriously expects that the $X$ particle
has spin two. Nevertheless, this is the only available `straw person' with which to compare
the spin-zero expectation, in the same spirit as the angular distribution of three-jet events in $e^+e^-$
annihilation were calculated long ago for the (unexpected) scalar gluon case~\cite{EK}, to be compared with the distribution for
the (confidently expected) vector gluon case. That comparison subsequently provided the first
experimental verification that gluons indeed have spin one~\cite{TASSO}, confirming the theoretical expectation.

Theorists expect the contest between spin zero and spin two to be like a match between
Brazil and Tonga~\footnote{{\tt http://resonaances.blogspot.com/2012/10/higgs-new-deal.html}}. 
The question is: what is the game - football (soccer) or rugby?

\section*{Acknowledgements}

VS would like to thank Eduard Masso for discussions on custodial symmetry. The work of JE was supported partly by the London
Centre for Terauniverse Studies (LCTS), using funding from the European
Research Council via the Advanced Investigator Grant 267352.
The work of TY was supported by a Graduate Teaching Assistantship from
King's College London. JE and VS thank CERN for kind hospitality.

\end{document}